\documentclass{PoS}
\usepackage{epsfig}
\usepackage{graphicx}
\usepackage{longtable}

\newcommand{\solm}{M$_{\odot}$\ }

\title{Experimental Indicators of Accretion Processes in Active Galactic Nuclei}

\ShortTitle{Accretion Indicators}

\author{\speaker{A. Eckart$^{~1\footnote{email: eckart@ph1.uni-koeln.de}~,2}$},
M. Valencia-S.$^{~1}$,
B. Shahzamanian$^{~1}$, M. Zajacek$^{~2,1}$, L. Moser$^{~1,3}$, 
G. Busch$^{~1}$, M. Parsa$^{~2,1}$,
M. Subroweit$^{~1}$, F. Peissker$^{~1}$, N. Sabha$^{~1}$, S.E. Hosseini$^{~1,2}$,
M. Horrobin$^{~1}$, C. Straubmeier$^{~1}$,
N. Fazeli$^{~1}$,
A. Borkar$^{~4}$, 
D. Kunneriath$^{~4,6}$, V. Karas$^{~4}$,
C. Rauch$^{~2}$, S. Britzen$^{~2}$, A. Zensus$^{~2}$,
M. Garc\'{i}a-Mar\'{i}n$^{~5}$, Y.E. Rashed$^{~7}$
\\
1) I. Physikalisches Institut der Universit\"at zu K\"oln, Z\"ulpicher Str. 77, D-50937 K\"oln, Germany;
\\
2) Max-Planck-Institut f\"ur Radioastronomie, Auf dem H\"ugel 69, D-53121 Bonn, Germany;
\\
3) Argelander-Institut f\"ur Astronomie, Universit\"at Bonn, Auf dem H\"ugel 71, 53121 Bonn, Germany
\\
4) Astronomical Institute of the Czech Academy of Sciences, Bocni II 1401, CZ-14100 Prague, Czech Republic
\\
5) European Space Agency (ESA/STScI), 3700 San Martin Drive, 
Baltimore, MD 21218, USA
\\
6) North American ALMA Science Centre, NRAO, Edgemont Road, Charlottesville, VA 22903, USA
\\
7) Department of Astronomy, Faculty of Science, 
University of Baghdad, 10071 Baghdad - Aljadirya, Iraq
}

\abstract{
Bright Active Galactic Nuclei are powered by accretion of mass onto the
super massive black holes at the centers of the host galaxies. 
For fainter objects star formation may significantly contribute to the luminosity.
We summarize 
experimental indicators of the accretion processes in Active Galactic Nuclei (AGN), i.e., 
observable activity indicators that allow us to conclude on the nature of accretion.
The Galactic Center is the closest galactic nucleus that can be studied with unprecedented
angular resolution and sensitivity. 
Therefore, here we also include the presentation of recent observational results on Sagittarius~A* 
and the conditions for star formation in the central stellar cluster.
We cover results across the electromagnetic spectrum 
and find that the Sagittarius~A* (SgrA*) system 
is well ordered with respect to its geometrical orientation and its 
emission processes of which we assume to reflect the 
accretion process onto the super massive black hole.

}

\FullConference{Accretion Processes in Cosmic Sources - APCS2016 -\\
		5-10 September 2016,\\
		Saint Petersburg, Russia}

\begin{document}

\section{Introduction}

Active Galactic Nuclei (AGN) are powered by accretion of mass towards their 
nuclei. The luminous AGN predominantly with cores that have 10 to 100 times
the luminosity of the entire host galaxy, gain their power through accretion 
onto super massive black holes (SMBH 10$^6$ to 10$^{10}$ \solm).
Massive and large scale star formation may also consume a large portion of
the matter and drive a considerable portion of the AGN luminosity.
Accretion of matter onto very massive and compact objects represents 
a very efficient conversion of potential and kinetic energy into radiation.
Hence the main power is due to the strong gravitational field that can 
be attributed to these compact objects, i.e. SMBHs.

Investigations of secondary phenomena like stellar dynamics, nuclear 
winds and jets, now lead to the assumption  that SMBH  exist in the 
centers of most massive galaxies.
The mass of the super massive black hole correlates  with 
the velocity dispersion of the galactic bulge 
(this is known as the the M-$\sigma$ relation; 
Ferrarese \& Merritt 2000; Gebhardt et al. 2000; Tremaine et al. 2002
or with bulge luminosity.
Central massive black holes may even be present at the centers of all 
galaxies down to the mass scale of globular clusters (L\"utzgendorf et al. 2013).

It is, however, not clear what the interplay between various mechanisms is
that lead to accretion onto the SMBH or prevent matter being accreted onto 
these massive objects.
While the immediate SMBH feeding process involves the presence of 
some form of an accretion disk (or at least accretion zone), 
the matter first has to loose angular momentum in a significant way and has
to get into this region.
The accretion disk itself and the immediate surroundings of the SMBH
have characteristic spectral signatures that allow us to study the 
accretion process - supported by theory and observations.

\begin{figure}[!ht]
\begin{center}
\includegraphics[width=13cm]{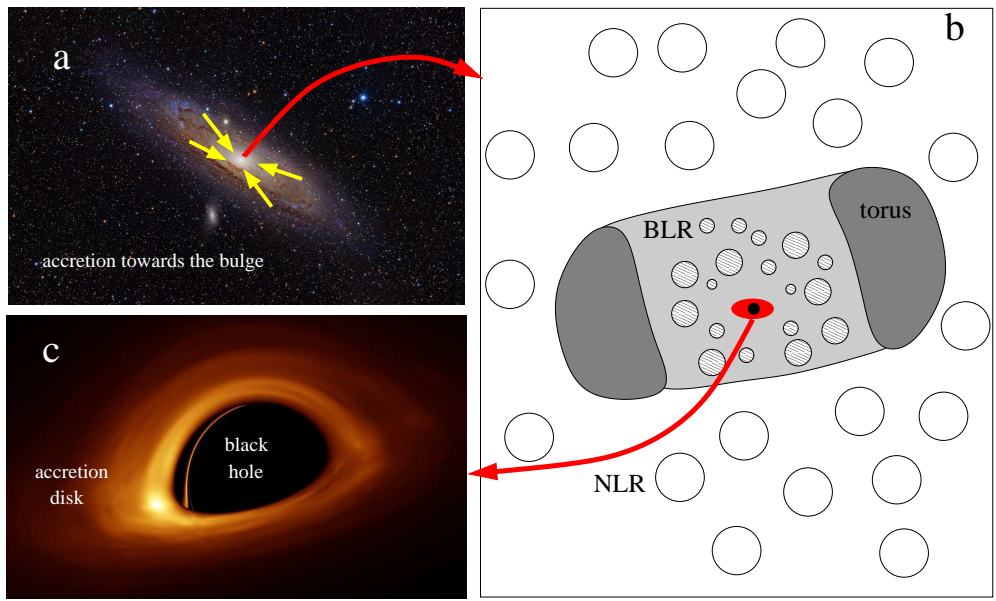}
\caption{
Overview of the accretion phenomenon on galactic scales. Zooming into the center of the 
galactic bulge with a gradually increasing resolution: 
(a) Gas and dust proceed in the inward together with sinking stars. 
The galactic bulge mass is correlated with the super massive black hole (SMBH) mass  
residing in the core of the galaxy (Magorrian et al. 1998); 
(b) On a few parsec scale, a dusty/gaseous torus encircles SMBH, together with 
clouds of Broad and Narrow Line Regions (BLR/NLR). Here, also stars of the dense 
Nuclear Star Cluster influence the environment (Urry \& Padovani 1995); 
(c) the innermost region contains a hot accretion flow extending down to 
its inner rim near above the Innermost Stable Circular Orbit (ISCO). 
The image and the spectrum are distorted by strong-gravity effects of the 
central black hole, which also define the shortest time-scale of the 
source variability (Bursa et al. 2007).
} \label{fig0}
\end{center}
\end{figure}

In Fig.~\ref{fig0} we give a basic overview over the general process
of accretion towards the center.
While variations in the potential field (mergers, bar instabilities, 
tidal effects; e.g.  Capelo \& Dotti 2017)
as well as viscosity effects allow matter to move towards the central region 
(e.g. Blank \& Duschl 2016; top left upper panel in Fig.~\ref{fig0}),
only a part of it reached the very nucleus and is then found in the 
typical AGN arrangement of Narrow Line Region (NRL), 
Broad Line Region (BLR), torus accretion disk and super massive black hole
(right panel in Fig.~\ref{fig0}).
It must be mentioned that the mass flow towards the central region of a host
is not necessarily directed towards the center at all times - the dynamical
arrangement may also be such that the matter is moving away from 
the nucleus for at least a fraction of the typical dynamical time scale.
Molecular gas torque maps of 25 galaxies of the NUGA sample reveal that only
1/3 of the systems shows indications for gas moving towards the center
on the several 100pc to kpc scales (Garcia-Burillo \& Combes 2012;
see also Smajic et al. 2015, Busch et al. 2017).
Finally a fraction of the matter reaches the accretion disk and 
can be accreted onto the SMBH due to viscous processes 
(left lower panel in Fig.~\ref{fig0}; and e.g. Blank \& Duschl 2016; Collin \&  Zahn 2008).

Since the SMBHs are compact one needs high angular resolution techniques to
study the details of the accretion processes or - at least - to 
clearly separate the AGN emission from the other emission sources - like
star formation. This is achieved best,  if the object of investigation is 
close to the observer. This implies the the nucleus of our Galaxy,
Sagittarius~A*, is the ideal target to investigate the physics of the 
accretion process. 

Sagittarius~A* (Sgr~A*) at the center of our Galaxy is a highly variable 
near-infrared (NIR) and X-ray source which is associated with a 
$4 \times 10^{6}$ \solm super massive central black hole (SMBH).
This region allows us in an unprecedented way to study at the 
same time, the nuclear activity associated with accretion onto 
the super massive black hole, and the possibility for 
star formation in its vicinity
How to form stars with a standard Salpeter mass function near a SMBHs is shown by Jalali et al. (2014).
Predominantly high mass star formation has been targeted by Nayakshin et al. (2005, 2007).
For observational evidence for star formation near SgrA* see, e.g., 
Eckart et al. (2013, 2004a) and Yusef-Zadeh et al. (2015, 2016).
Accretion and star formation models must take into account 
the presence (or influence) of
several dusty sources in the central 1~pc region
(Meyer et al. 2014, Eckart et al. 2013).
The fast motion of one of these infrared excess sources 
was discovered by Gillessen et al. (2012).
They interpret it as a core-less (i.e. no star at its center) 
gas and dust cloud approaching SgrA* on an elliptical orbit.
Eckart et al. (2013) present K$_s$-band identifications 
(from VLT and Keck data) and proper motions of this 
Dusty S-cluster Object (DSO; also called G2 in the literature).
The S-cluster is the cluster of high velocity 
stars surrounding SgrA* (see Eckart \& Genzel 1997). 
The DSO passed by SgrA*
within $\sim$150-160~AU in spring 2014..
In this contribution we present a brief description of the 
accretion process acting onto SgrA*
from a statistical analysis of emission at different wavelengths. 
We also summarize the exploration of the geometrical properties 
of SgrA* accretion flow and of the DSO via infrared polarimetry
(Shahzamanian et al. 2015, 2016).

\section{Star formation and Black Hole Growth}

\begin{figure}[!ht]
\begin{center}
\includegraphics[width=13cm]{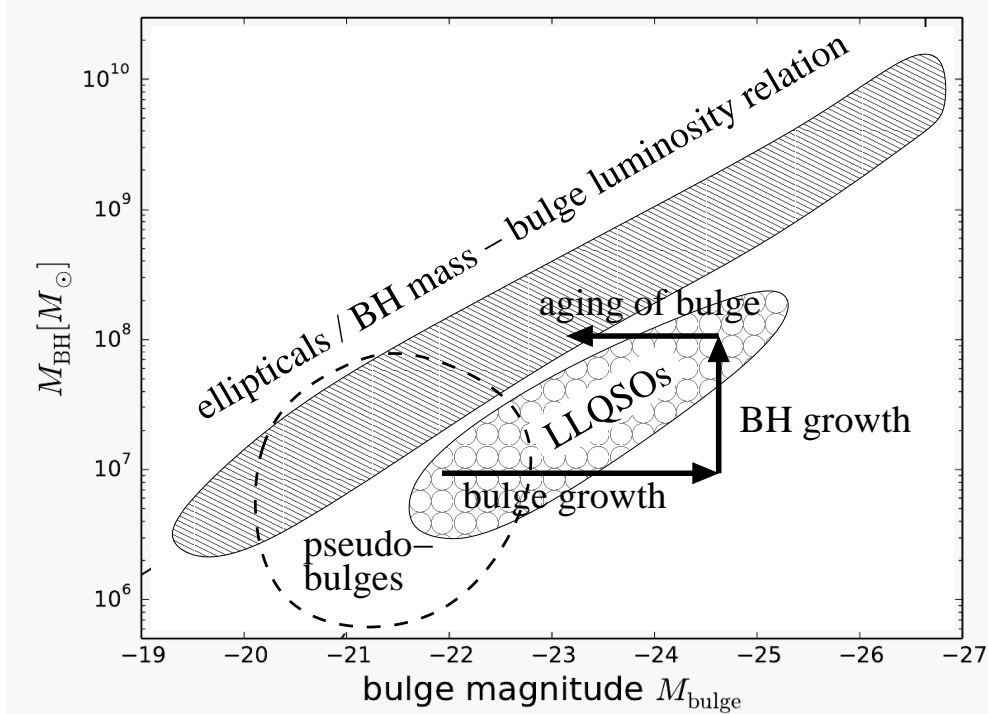}
\caption{
A possible evolutionary scenario in the black hole mass - bulge luminosity diagram.
Accretion of matter onto the central region results into enhanced star formation and black hole growth.
Young stellar populations cause 
over-luminous bulges compared to inactive galaxies on the relation. 
Black hole growth and aging of the stellar populations then move the objects back onto the relation.
} \label{fig1}
\end{center}
\end{figure}

There are strong indications that the mass of the central super-massive black holes are linked to 
properties of their host galaxies. 
Here, understanding the properties of the host galaxies of quasi-stellar objects (QSOs) is important 
in order to study possible co-evolution of both components.
These studies can best be pursued for Low-luminosity type-1 QSOs (LLQSOs) as these objects are 
cosmological very close and hence, allow for a detailed analysis of their host galaxy.
In Busch et al. (2014) we selected a small sample of the Hamburg/ESO
survey for bright UV-excess QSOs consisting of 99 Low Luminosity QSOs with redshift z$\le$0.06. 
We performed near-infrared J-, H-, and K-band imaging of 20 host galaxies
of this subsample.
We then used the reduced images to carry out an aperture photometry and bulge-disk-decomposition.
We separated the disk, bulge, bar, and nuclear components. 
More than 40\% of these sources turned out to be barred disk galaxies and 
about 50\% were elliptical or $S0$ galaxies.
The stellar masses span an interval from about 2$\times$10$^9$\solm to 2$\times$ 10$^{11}$\solm.
The average mass amounts to 7$\times$10$^{10}$\solm. The spectroscopically derived black hole 
masses cover an interval from  10$^6$\solm  to 5$\times$10$^8$\solm with a 
median mass of 3$\times$10$^7$\solm. 
These objects populate a region between the Seyferts and QSOs that  has not been well studied so far. 
For 16 LLQSO, we have BH masses and bulge luminosities at hand, which 
enables us to study their location in the BH mass vs. bulge luminosity 
relation. We find that all of them lie below published relations for 
inactive galaxies (Busch+16, Fig.~\ref{fig1})."
This points at possible evolutionary processes between 
the host bulge and the SMBH.
Two possibilities can be discussed: The first possibility is that the bulges of active galaxies contain
much younger stellar populations in comparison to the bulges of inactive galaxies. 
In a bulge luminosity vs. black hole mass diagram this moves the data points toward higher luminosities.
A second possibility is that the black holes of these AGN are under-massive compared to hosts of 
similar mass and luminosity.
It might indicate that the black holes are still accreting mass in a substantial way and that the
mass-assembly of the host spheroid is occurring with a faster rate.
Further black hole mass accretion would move the sources toward higher black hole 
masses in the corresponding diagram.

However, the central regions of many low-mass galaxies tend to be  more disky with 
younger stellar populations at their centers which then are less similar to classical bulges. 
These are called 'pseudobulges'. Being more disky implies higher rotation and lower velocity dispersion 
towards the center - resulting in a lower apparent bulge mass
(Kormendy and Kennicutt, 2004, Kormendy et al. 2011). 
It is unclear in how far this effect needs to be considered in the 
case of LLQSOs.
Our study Busch et al. (2015) suggests that only near-infrared 
integral-field spectroscopy with adaptive optics gives us the 
necessary resolution to study what is happening in the central 
bulges of these sources. We show that at least in the example case 
of HE1029-1831 the deviation from BH mass - bulge luminosity 
relations of inactive galaxies is caused by overluminosity due 
to intermediate-age stellar populations in the central region."

A strong signature of black hole accretion is AGN activity that 
may be enhanced due to merger processes.
Merging of galaxies and - as a consequence - the merging of the two central super massive black holes 
appears to have been more important in the past at redshifts of z=1-2 of higher.
Recent observations of AGN in galaxies with stellar masses above 
10$^{10.4}$\solm have shown that at lower redshifts AGN hosts do 
not show enhanced merger signatures if they are compared with normal galaxies, 
this means that most of the low redshift AGN are hosted by quiescent 
host galaxies. 
At redshifts above z=1 the AGN are found more often in star-forming galaxies.
Micic et al. (2016) show that the AGN occupation fraction in star-forming and 
quiescent hosts follows the evolution of the galaxy merger rate that drops strongly at z$<$1. 
At these redshifts only old mergers trigger AGN activity  and most active sources are 
in their declining phase.

\section{Relativistic radio jets}

Jet activity is an indication of accretion onto the central system consisting of a SMBH and an accretion disk.
Matter lost through the jet or entrained with it is not available for accretion anymore.
Jets are also thought to trigger star formation and to be a significant component in feedback  
between the AGN and inter cluster media or the star forming ISM (Yang \&  Reynolds 2016; Rashed et al. 2013).
Jets are very likely an indicator for the way accretion is taking place close to their foot points
(see, e.g., Britzen et al. 2017ab).
Corkscrew shaped jets may indicate either a binary black hole or at least some disturbing mechanism
(possibly periodic or quasi-periodic) that influences the launching mechanism, i.e., the accretion disk
(Britzen et al. 2015; Sobacchi, Mattia \& Stamerra 2017).

Lister et al. (2016) present 1625 new 15 GHz (2 cm) VLBA maps of 295 Parsec-scale jets with 
than  that are associated with  AGN The sources were part of the MOJAVE and 2 cm VLBA surveys that 
were collected between 1994 and 2013.  The jets contain several 100 components,
many of which could be traced 
kinematically over at least 10 different observing epochs.
More than  half of these showed non-radial and/or accelerating motions indicating the presence of complex
non-ballistic motions in jets. 
The apparent velocities peak around 5 times the speed of light, however, a few jets have speeds of 
up to 30c.
These speeds indicate large Lorentz factors an physical speeds close to the speed of light.
A comparison between $\gamma$-ray activity and super luminal motions suggests that sources 
with high $\gamma$-ray fluxes also have on average high velocity jets.
A number of Seyfert~1 sources also show jet activity comparable to higher luminosity BL~Lac 
(BL~Lacetrae)
and flat spectrum quasi stellar radio sources (quasars).

However, the physical nature of the jet-components as well as the launching mechanism are largely unknown. 
In Britzen et al. (2017a), we present an interesting example: the jet in the quasar 1308+326 at z=0.997.
The MOJAVE (Monitoring Of Jets in Active galactic nuclei with VLBA Experiments) survey allowed for a longterm
tracking of multiple jet components.
Individual jet components could be traced over about two decades making use of a total of 49 
observing periods at 15~GHz.
Additional experiments provided a frequency coverage between 4.8, 8.0, and 14.5 GHz in polarization 
and total power.
Detailed modeling suggests that the jet features are launched under varying viewing angles 
into an ejection cone apparently also under the influence of some rotational motion in addition 
to the motion along the jet resulting into a helical jet structure.
The observations can be explained by plasma moving along magnetic filed lines of magnetic flux tubes.

Britzen et al. (2010) 
present evidence that parsec-scale jets in BL~Lac objects may be significantly distinct in 
kinematics from their counterparts in quasars. We argued this previously for the BL~Lac 
sources 1803+784 and 0716+714 and report here a similar pattern for another well-known BL~Lac object, 
PKS 0735+178, whose nuclear jet is found to exhibit kinematics atypical of quasars.
However, the classic paradigm of apparent superluminal motion is not always an optimum 
description of the jet kinematics.  In the case of the BL~Lac object 0735+178, Britzen et al. (2010) 
found a significant structural mode change. In the course of 1-2 years the jet trajectory 
was bent twice sharply and made a transition from 'typical superluminal' into a 
quasi 'stationary' state with linear jet shape. 
Part of the changes may also be correlated with variations of the viewing angle of difference knots,
however, several model changes could be traced back and seem to be correlated with optical 
and/or radio outbursts.
Such a behavior was also found for the BL~Lac objects 1803+784 and 0716+714
and may be a key to understand the difference between quasars and BL~Lac objects.

\section{Reverberation: response to nuclear variability}

The reverberation mechanism represents the immediate answer of 
the ionized gas surrounding the SMBH to the continuum variability of the accretion flow or disk.
Not only does the mechanism allow us to probe this region and determine the
sizes and extend of the reverberation matter or the hardness of the ionizing spectrum,
but is reassures us at the same
time that the variable radiation is indeed coming from the very nucleus
of the corresponding source and not from secondary structures like outlying
jet components or interactions of the jet or outflow with the interstellar
medium within the galaxy.
We find reverberation signals from both: the BLR and the NLR
revealing information on different size and time scales.

\subsection{NLR reverberation: response to long term variability}

Using the MODS spectrograph at the Large Binocular Telescope (LBT), Rashed et al. (2015)
compared optical spectroscopic and photometric data for 18 AGN
galaxies over 2 to 3 epochs, spanning over a time interval of typically 5 to 10 years.
The spectra were compared with SDSS and literature data. 
Rashed et al. (2015) report significant  variations in the forbidden oxygen and the 
hydrogen recombination lines of several sources.
The authors show that it is most likely the 
difference in black hole mass between the stronger (broad line spectrum) and less variable 
(narrow line spectrum) subsets of the sample 
that is responsible for the different strength in the continuum variability. 
In the following $\Delta$ denotes differences between the two samples.
Rashed et al. (2015) could show that for an approximately constant accretion rate the
total Narrow Line Region line variability reverberates the continuum variability.
The entire data set points at a relation between the line ($\Delta$L$_{line}$) and continuum
($\Delta$L$_{cont}$) variability of $\Delta$L$_{line}$$\propto$($\Delta$L$_{cont}$)$^{3/2}$.
Since this dependence (shown by 61 BLR and NLR sources from the literature) 
is also strongly found in in the narrow line emission, this also implys that the 
luminous part of the NLR must be very compact, with a diameter in the range of 10 light-years.
A simlar NLR reverberation response has been reported for two other sources:
the radio galaxy 3C390.3 (Clavel \& Wamsteker 1987) and for the S1 galaxy NGC 5548 (Peterson et al. 2013).

While it is known tha the low luminosity part of the NLR can be quite extended such that the NLRs
in many nearby sources can actually be spatially resolved, 
it is more difficult to find reliable information on the size of the high surface luminosity part of the NLR.
The density and temperature gradient towards indicated an enhancemant of both quantities
towards the center (Ogle et al. 2000; Penston et al. 1990; Morse et al. 1995), leading to the fact that
the surface brightness distribution of the NLR flux is very centrally peaked. 
Bennert et al. (2006a,b) 
found from HST scans across seyfert nuclei that more than half the line flux 
orignates from within the central 10-20 parsecs.

\begin{figure}[!ht]
\begin{center}
\includegraphics[width=13cm]{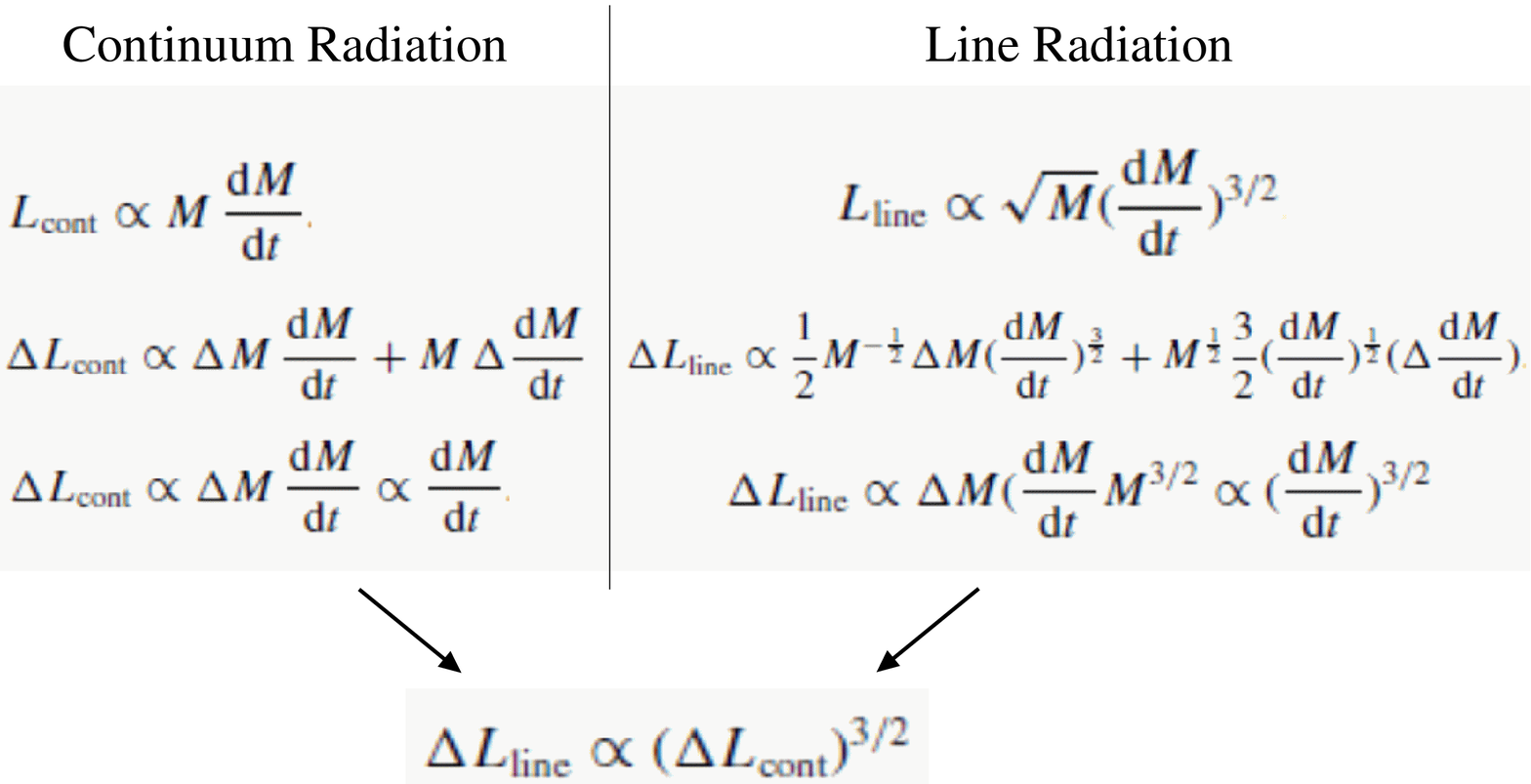}
\caption{
Formalism that leads to the relation between the degree of continuum and line variability.
$\Delta$L$_{line}$$\propto$($\Delta$L$_{cont}$)$^{3/2}$.
See text and Rashed et al. (2015) for details.
We used that the continuum and line luminosity is proportional to the 
accretion rate and the 
luminosity of the nucleus, respectively.
On the left hand side we assume an approximately constant accretion rate.
On the right hand side we use that the ratio between surface and the volume is proportional 
to the square root of the continuum luminosity.
The NLR is large but very compact, i.e. the brightness is centrally peaked.
Here, $\frac{dM}{dt}$ are the accretion rates and
$\Delta M$ is the median SMBH mass difference between dominantly narrow and broad line spectrum
dominated sources of the two samples treated by Rashed et al. (2015).
} \label{fig2}
\end{center}
\end{figure}

\subsection{BLR reverberation: short term response: BLR/size/map}

The broad emission line regions in AGNs are dominated by
photoionizations from the central source.
Over short time intervals of hours to days reverberation mapping 
(see reviews by, e.g., Denney et al. 2016, Peterson 1993, Horne et al. 2004)
allows us to investigate the structure of the BLR.
As a valuable result it provides an estimate for the size of the BLR that can recombined with the BLR velocity
dispersion of the broad lines ($\sigma_{BLR}$) from spectroscopy to get an estimate of the black hole mass.
The size $R_{BLR}$ of the BLR is too small to be imaged by current instrumentation. 
Their sizes are light hours to light days - depending on the ionization potential of the tracer ion.

The gas mass of the BLR is small compared to that of the NLR.
At BLR radii the amount of 'contaminating' stellar mass is minimal.
Hence, the black hole masses 
derived via this method belong the the most reliable estimates.
The black hole mass $M_{BH}$ is given by:

\begin{equation}
M_{BH} \sim f \times R_{BLR} (\sigma_{BLR})^2~~~.
\end{equation}
	
The factor $f$ reflects the exact shape of the BLR.
The biggest difficulty with applying this formula is the measurement of the BLR radius. 
One standard technique is based on the 
fact that the emission-line fluxes vary 
strongly in direct response to the changes in the underlying continuum, i.e., the light from the 
accretion disk close to the black hole. Hence, the expression 'reverberation'. 
One finds that the emission-line response is delayed with respect 
to variations in the nuclear continuum. If the delay is due to the light travel times, 
it reflects the size of the emission-line region.
Dependencies of the size on the ionization potential of the tracer ions are contained in the 
constant 'f' and makes a straightforward 
3-dimensional mapping impossible but allows size measurements.
There are now several tens of AGNs and quasars (most are at z < 0.3) 
for which measurements of the average time lag are available (e.g., Kaspi et al. 2000, 
Peterson et al. 2004, Bentz et al. 2009b).

\subsection{Variability and time lags: accretion disk size and structure}

In general the degree of variability increases with source type along the
Seyfert, QSO/quasar, BL~Lac, BLASAR/optically violent variable sequence.
For many of these sources one can obtain information of the accretion
disk structure.  
For a few sources relativistic Fe K$\alpha$ line profiles can be observed.
These profiles results from material at high speed close 
close to the strong gravitational well of the central super massive compact object
(e.g., Liu et al. 2016; Marin \& Tamborra 2014).
Most likely the material is orbiting the SMBH close to the last stable orbit.
Independent of its arrangement and origin, emission from this region is a clear 
indication, that matter is or will be accreted onto the central object.

In general the variability is strongly influenced by or actually reflects 
variations in the accretion and - at least in the case of radio activity -
the line of sight to the observer through beaming effects.
For the Seyfert 1 galaxy NGC 5548, Edelson et al. (2015)
recently densely monitored several UV/optical 
bands and the X-rays over a time range of more than 100 days with a
mean sampling rate better than half a day.
The UV/optical light curves show time lags the obey a 
$\approx \lambda^{4/3}$ wavelength dependence.
If the time lags are interpreted as origination from the accretion disk 
it results in a rather large size of the order of 1/3 light day.
Combining this size measurement with those obtained from quasar 
micro-lensing studies implies that 
over a wide range of black hole masses the AGN disk sizes approximately scale 
linearly with the mass of the central black hole.

Radio-loud active galactic nuclei (AGNs) consist of a 
super massive black hole (SMBH) in the mass range of 10$^6$ to 10$^9$ \solm
and a relativistic jet emitting highly polarized synchrotron radiation.
Several of these AGN show fast flux variability on a timescale of  a few hours
to a day. This type of variability is referred to as intra-day variability 
(IDV; Heeschen et al. 1987). 
IDV is found from the radio cm- to mm wavelength range to the optical domain
(Quirrenbach et al. 1991; Wagner et al. 1996; Kraus et al. 2003; Gupta et al. 2008).
At long cm radio wavelengths this variations are mainly caused by extrinsic reasons 
like interstellar scintillation (Rickett 1998).
The jet is moving close to the speed of light.
Therefore, at short radio wavelengths to the optical domain intrinsic source 
variations from the jet are magnified by the relativistic beaming (Wagner \& Witzel 1995). 
The intrinsic variations are usually interpreted as being caused by
the propagation of shocks in the jet (Marscher \& Gear 1985).
The S5 source 0716+71 is well know for its strong IDV.
Lee et al. (2016) studied  millimeter-wavelength variability 
of the polarized emission of S5 0716+714 at  22, 43, and 86 GHz
using the Korean Very Large Baseline (VLBI) Network. 
Over a period of 10 hours this source shows significant variations 
in the polarization angle (several 10 degrees) and degree of polarization (several percent). 
The data indicates that the polarized emission from S5 0716+714 
is transferred through a Faraday screen located in or near the jet of the source.

\section{SgrA* as a special nearby case}

The SMBH at the center of the Milky Way, SgrA*, can be studied with the
highest angular resolution across the electromagnetic spectrum.
Although, as an exceedingly weak nucleus, its emission can clearly 
be separated from the surrounding stars in the NIR.
In the radio to X-ray domain it also can clearly be distinguished from the 
emission of the gas and dust components close to the line of sight.

\subsection{Radio/sub-mm single dish and VLBA monitoring}

SgrA* is strongly variable at all wavelengths.
Most of this variable flux density arises from the innermost region of the accretion
flow onto the central super massive black hole (e.g. Moscibrodzka et al. 2009, 2013).
A detailed statistical analysis of the 345~GHz sub-millimeter and the 100~GHz radio 
flux density variation is presented by Subroweit et al. (2017) and Borkar et al. (2016).
SgrA* shows flux variations of 0.5 to 1.0 Jy over time scales of typically 1.5 to 3 hours. 
The millimeter flux density distribution and the variability of SgrA*
were studied in detail by
Moser et al. (2016) using the 
Atacama Large Millimeter/submillimeter Array (ALMA).
At a resolution of up to 0.5'' the investigation was carried out over the central parsec
surrounding the SMBH SgrA*. 
These observations resulted in 
serendipitous detections of line emission
over the mm-wavelength range
resulting in detailed temperature and density 
estimates towards different source components in the central arcminute of the Galaxy.

The observations clearly revealed a varying spectral index (S$\propto$$\nu$$^{\alpha}$) 
of the SgrA* synchrotron emission.
In a frequency range of 100 - 250~GHz the spectral index is $\alpha$=0.5 and in the 
230 - 340~GHz interval Moser et al (2016) find $\alpha$$\sim$0.0.
Previous interferometric and single dish sub-mm observations 
as well as theoretical modeling have shown an
overall spectral index drop towards 
frequencies higher than $\sim$350 GHz 
(Marrone 2006,  Marrone et al. 2006a,b, Eckart 2012a).
In this frequency domain other Galactic Center regions (e.g. the mini-spiral or the CND;
Kunneriath et al. 20012, Garcia-Marin et al 2011)
show continuum emission 
from Bremsstrahlung (around $\alpha$=-0.1) 
as well as emission from cold dust with a flat or even inverted spectrum.
The investigation by Subroweit et al. (2016) and Borkar et al. (2016)
has shown that the 
variable millimeter emission of SgrA* can be explained 
via an adiabatically expanding plasmon model with a peak frequency 
around 345~GHz.
The data indicate that the expanding source components 
are either confined in the immediate vicinity of the SMBH or exhibit a bulk motion greater 
than their adiabatic expansion velocity.
It is unclear were these components are being formed.
This happens either in the temporary disk or in the corona of the SMBH SgrA*.
When the source components are most compact they result in variable 
thin NIR emission. The strongest variations produce variable X-ray emission via the Synchrotron Self-Compton (SSC) effect.

During the flyby of the dusty S-cluster object (DSO)
the flux-density variations were not obviously affected.
In agreement with infrared line and comtiuum observations (Valencia et al. 2015)
this demonstrates that the source is very compact rather than an extended gas and dust source.

\subsection{Variable radio structure and the stability of the SgrA* system}

Near-infrared triggered mm-VLBI observations of the radio source SgrA* at 43~GHz
(Rauch et al. 2016)
allow us to obtain information on the 
nature of some of the SgrA* (sub-)millimeter flares.
Monitoring the 7 mm (43~GHz) flux density with the a global Very Long Baseline Array (VLBA) campaign
over 6 hours each day we found only small ($\le$0.06~Jy) variations of the total flux.
Through a NIR flare observed with the  Very Large Telescope (VLT), however, we triggered, however, 
VLBA observations. The NIR flare preceeded a 43~GHz flare of 0.22 Jy by about 4.5 hours.
Such a time delay is a typically expected value in case of adiabatic expansion. 
Close to the maximum of the mm-flare, SgrA*
showed a secondary radio component at about 1.5 mas distance southeast of the SgrA* core position. 
One would expect such a component at this distance in case of high
bulk relativistic motions ($>0.5~c$) of the expanding source component.  
as it is found in a jet or an outflow component.
(see e.g.  Yusef-Zadeh et al. 2006, Yuan et al. 2009, Li et al. 2017).
Such changes in the intrinsic source structure of SgrA* have been reported 
in several cases (see, e.g., Bower et al. 2014).
However, it cannot be excluded that scintillations of the radiation within the plasma
on the line of sight to SgrA* are contribution to structural variations as well
(Ortiz-Leon et al. 2016).

At NIR wavelengths SgrA* is strongly linearly polarized
with typical polarization degrees around 20\%
and a polarization angle centered around 13$^\circ$$\pm$15$^\circ$.
This allows us to derive information on both the emission process and the overall geometry 
of the emitting structure.
For high polarized flux densities the flare flux statistics is consistent
with the single state power-law distribution derived by Witzel et al. (2012).
Details are given by Shahzamanian et al. (2015).
All these facts show that the source geometry and the energetics of the accretion process 
within the SgrA* system are quite stable.
The SgrA* system consists of a black hole, a wind or jet, and a temporary disk or mid-plane accretion flow as 
found in simulations (e.g. Moscibrodzka et al. 2009, 2013).

\subsection{X-ray activity of SgrA*}

For its mass SgrA* is sub-luminous in the X-ray domain.
We only measure 
a bolometric luminosity of $L_{bol}\sim~10^{36}~erg/s$.
This is well below the
Eddington luminosity of 
$L_{Edd} = 3 \times 10^{44} erg/s$
(Yuan et al. 2003).
Most likely the low luminosity of SgrA* can be explained by radiatively inefficient accretion 
flow models (such as advection dominated accretion flows - ADAF; 
Narayan et al. 1998). Alternative models involve a jet-disk structure.
However, SgrA* exhibits strong X-ray flares -
 - first found by Baganoff et al. (2001).
Detailed monitoring has revealed that one finds about one brighter flare per day (Neilsen et al. 2013) 
that accounts approximately 10 times the quiescent 2-8~keV flux of SgrA*.
This quiescent flux is due to an extended source of Bremsstrahlung emission surrounding 
the SMBH and accounting for about
$3.6 \times 10^{33} erg/s$
(Baganoff et al. 2003; Nowak et al. 2012).
Extreme flares with brightnesses exceeding 150 times this quiescent level have been 
measured  and reported by  Porquet et al. (2003, 2008) and Nowak et al. (2012). 

The flare X-ray flare emission is consistent with SSC scattered radiation from 
synchrotron components 
the optically thin  radiation of which, can be observed in the NIR.
Here the spectral index of the variable flux density is
about $\alpha$=-0.7 ($S_{\nu} \propto \nu^{\alpha}$).
The statistical properties of the flare amplitude have been described by Witzel et al. (2012).
The SSC mechanism needs the least demanding requirements for the production of X-ray luminosity.
Only moderate densities of relativistic electrons  of ($10^6 cm^{-3}$)
with $\gamma$$\sim 10^3$ are required
(e.g. Eckart et al. 2004, 2008, Yusef-Zadeh et al. 2006).
Detailed X-ray monitoring e.g. by Mossoux et al. (2016) has also revealed that the
X-ray luminosity did not change significantly during and past the flyby of the DSO.
Mossoux et al. (2016) have observed a total of seven NIR flares. Three of these
flares X-ray counterparts that were detected with XMM-Newton.
The flaring statistics was identical to the one obtained during the
2012 Chandra campaign.

\subsection{Monitoring the Dusty S-cluster Object:}
\label{section5}

In Valencia et al. (2015) and Peissker et al. (2017)  we report on our monitoring 
of SgrA* and the Dusty S-Cluster Object (DSO) in the
NIR domain using VLT SINFONI in the years between 2005 and 2015.
The faint DSO is orbiting SgrA* on a highly elliptical orbit with a periapse in May 2014.
The object is well tracable in the thermal NIR L'-band and in its Br$\gamma$-line 
emission in the NIR K-band.
Eckart et al. (2013, 2015) find prominent K-band continuum flux of the DSO.
The fain NIR K-band continuum is polarized.
A high degree of polarization indicates that the scattering material is surrounding the object in a 
non-spherical geometry (see, e.g., Zajacek, Karas \& Eckart 2014, Zajacek et al. 2016).
Shahzamanian et al. (2016) show that the polatization of the source is
consistant with its compact stellar nature.
The fact the the DSO stayed very compact in its continuum and line emission 
(Valencia et al. 2015,  Witzel et al. 2014) clearly demonstrates that it did not 
disintegrate during flyby of SgrA*  as it had been previously expected  
(e.g.  Gillessen et al. 2012, Pfuhl et al. 2015, Schartmann et al. 2015).

\begin{figure}[!ht]
\begin{center}
\includegraphics[width=13cm]{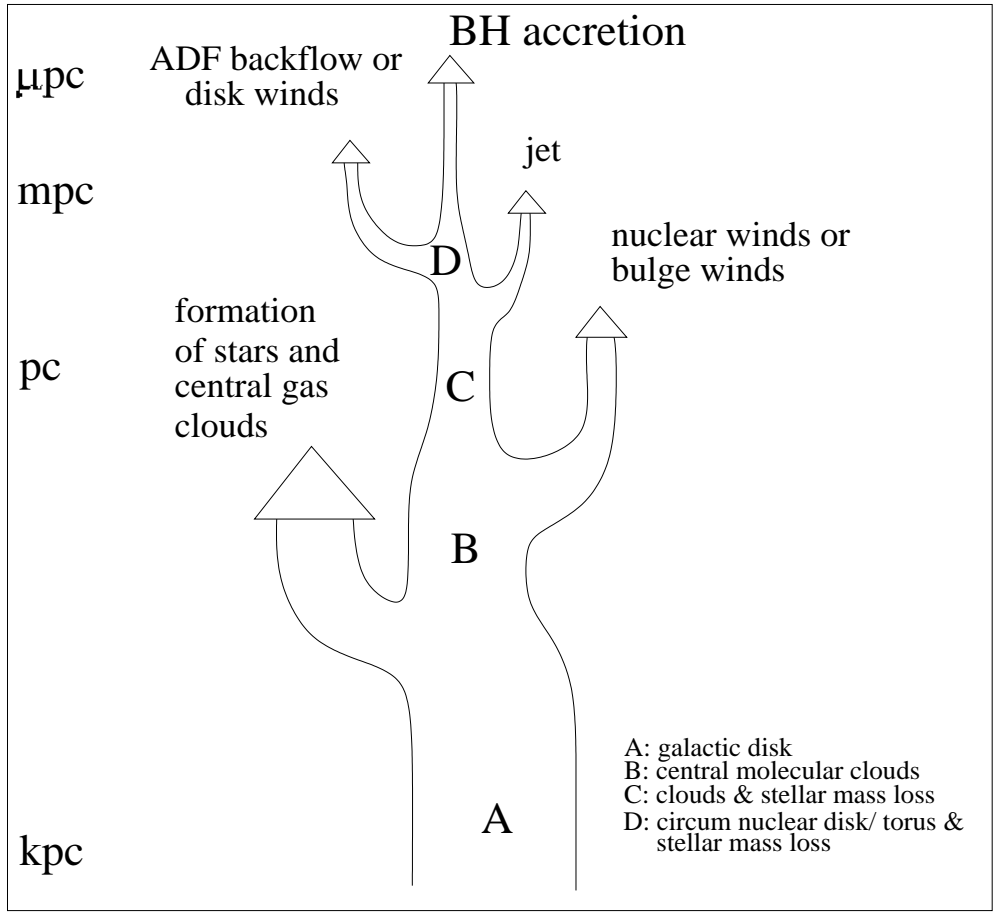}
\caption{
Accretion tree for super massive black holes.
Typical size scales are given on the left side.
} \label{fig3}
\end{center}
\end{figure}

\section{Summary}

We have essentials of processes that need to be considered if one
discusses the accretion of matter towards the nuclei of galaxies and
finally onto SMBHs.
The overview given  in 
Fig.~\ref{fig0} can also be presented in the form of an 'accretion tree',
as done in Fig.~\ref{fig3},
in which the interdependances between various mechanisms are highlighted 
as branching point of the tree.
The typical size scales decrease from bottom to top. 
The width indicates the amount of matter involved.
On kpc-scales
matter is transported towards the center via variations in the
potental field and viscosity.
The amount of matter that goes into star formation depends on the
dynamical state of the galaxy. If merger or bar formation processes are 
at work enhanced star formation might be the dominant process to
convert the centrally accreted matter in to luminosity.
Substantial amounts of matter may also be lost due to massive, global 
nuclear star formation winds that transport the material out of the 
system (as it is probably the case for star forming dwarf galaxies like M82)
or back to larger radii.
Closer to the AGN itself massloss may occur in the form of a jet or wind from
the immediate environment of the SMBH. For galaxies with low 
activity - as it is the case for our Milky Way, 
back winds from the Advection Dominanted Accretion Flow
(Narayan et al. 2012; Narayan \& Mahadevan 1998; Narayan, Garcia \& McClintock 1997), 
in which only a 
minute amount of matter reaches the SMBH, may occur. As in this case the 
accretion rate is a function of radius the dominant protion of the material
that could have been accreted is blown back to larger radii again.
For higher luminosities - and in case an accretion disk forms - irradiation
from the inner hot sections of the accretion disk onto sections of the disk
at larger radii may lead to an abrasion of the disk in the form of a mass 
loaded disk wind that leads to deep absorption troughs in the case of 
broad line absorbtion QSOs (Matthews et al. 2016, Higginbottom et al. 2013).
Finally a small amount of the matter is indead accreted onto the SMBH.
For inactive nuclei ,e.g. SgrA*, this accretion is 10$^9$ times smaller than
the Eddingtion rate (see SgrA* in comparison to LLAGN: Eckart et al. 2012b).
For SMBHs that are accreting heavily over their 
equatorial zone the accretion rate my reach values than exceed 
the Eddington accretion rate by far.

We have presented results of recent observations 
of SgrA*, the counterpart of the Milky Way's SMBH.
The low activity of SgrA* is characterized by flux density variations 
accross the electromagnetic spectrum.
A plausible mechanism is the formation and adiabatic expansion of 
synchrotron components. The  evolution of their spectra that are
peaked in the  submm-domain when they are born, can be studied in the 
mm- cm-radio domain. At lower radio frequencies a number of emission processes
from components of the overall accretion flow onto SgrA* are at work.
The optically thin part of the variabla synchrotron flux can be 
studies in the NIR domain.
At least the brighter flares are associated with X-ray flares through the 
SSC process.
Furthermore, studying the central cluster stars, one finds evidence 
for black-hole supported star formation, either via the formation 
of a massive star forming disk - the relicts of which are then blown away -
or by compression of individual dense molecular clumps  that may originate 
in the Circum Nuclear disk.
Several compact dusty sources including the Dusty S-cluster Object (DSO) 
may be in fact be reatively young and low-mass stars.

\section*{Acknowledgements}
We received funding
from the European Union Seventh Framework Program
(FP7/2013-2017) under grant agreement no 312789 - Strong
gravity: 'Probing Strong Gravity by Black Holes Across the
Range of Masses'.  
This work was supported in part by the
Deutsche Forschungsgemeinschaft (DFG) via the Cologne
Bonn Graduate School (BCGS), the Max Planck Society
through the International Max Planck Research School
(IMPRS) for Astronomy and Astrophysics, as well as special
funds through the University of Cologne and
SFB~956 - Conditions and Impact of Star Formation.,
and the Czech Science Foundation - DFG collaboration (No. 13-00070J).
M. Zajacek, M. Parsa and
B. Shahzamanian are members of the IMPRS.


\begin{thebibliography}{200}
\bibitem{} Baganoff, F. K., Bautz, M. W., Brandt, W. N., et al. 2001, Nature, 413, 45
\bibitem{} Baganoff, F. K., Maeda, Y., Morris, M., et al.  2003, ApJ, 591, 891
\bibitem{} Bennert N., Jungwiert B., Komossa S., Haas M., Chini R., 2006a, A\&A, 459, 55
\bibitem{} Bennert N., Jungwiert B., Komossa S., Haas M., Chini R., 2006b, A\&A, 456, 953
\bibitem{} Blank, M.; Duschl, W.J.; 2016, MNRAS 462, 2246
\bibitem{} Borkar, A.; Eckart, A.; Straubmeier, C.; Kunneriath, D.; et al., 2016, MNRAS 458, 2336
\bibitem{} Bower, G.C.; Markoff, S.; Brunthaler, A.; Law, C.; et al., 2014, ApJ 790, 1
\bibitem{} Britzen, S.; Witzel, A.; Gong, B. P.; Zhang, J. W.; Gopal-Krishna; et al., 2010, A\&A 515, 105
\bibitem{} Britzen, S.; Eckart, A.; L\"ammerzahl, C.; Roland, J.; 2015, AN 336, 471
\bibitem{} Britzen, S., Qian, S.-J., Qian, Steffen, W., Kun, E., et al., 2017a, A\&A 602, 29
\bibitem{} Britzen, S.; Fendt, C., Eckart, A.; Karas, V., et al. 2017b, A\&A 601, 52
\bibitem{} Bursa M., Abramowicz M. A., Karas V., et al., 2007, Proc. of Workshops on Black Holes 
       and Neutron Stars, eds. S. Hledik and Z. Stuchlik (Silesian University in Opava), pp. 21-25
\bibitem{} Busch, G.; Zuther, J.; Valencia-S., M.; et al., 2014, A\&A 561, 140
\bibitem{} Busch, G.; Smaji\'{c}, S.; Scharw\"achter, J.; Eckart, A.; et al., 2015, A\&A 575, 128
\bibitem{} Busch, G.; Fazeli, N.; Eckart, A.; et al., 2016, A\&A 587, 138
\bibitem{} Busch, G.; Eckart, A; Valencia-S., M; Fazeli, N; et al., 2017, A\&A 598, 55
\bibitem{} Capelo, P.R.; Dotti, M.; 2017, MNRAS 465, 2643; 
\bibitem{} Collin, S.; Zahn, J.-P.; 2008, A\&A 477, 419
\bibitem{} Clavel J., Wamsteker W., 1987, ApJ, 320, L9
\bibitem{} Denney, K.D.; Horne, K.; et al., 2016, ApJ 833, 33
\bibitem{} Eckart, A.; Britzen, S.; Valencia-S., M.; Straubmeier, C.; Zensus, J. A.; Karas, V.; 
        Kunneriath, D.; Alberdi, A.; Sabha, N.; Sch\"odel, R.; P\"utzfeld, D.	
	The Galactic Center Black Hole Laboratory, 2015arXiv150102171E, 2015
        in "Equations of Motion in Relativistic Gravity", D. Puetzfeld et. al. (eds.), 
        Fundamental theories of Physics, 179, pages 759-781, Springer 2015
\bibitem{} Eckart, A.; Horrobin, M.; Britzen, S.; Zamaninasab, M.; et al.;
           Proceedings of the International Astronomical Union, Volume 303, pp. 269-273, 05/2014
\bibitem{} Eckart, A., Muzic, K.; Yazici, S.; Sabha, N.; et al., 2013, A\&A 551, 18
\bibitem{} Eckart, A.; Garcia-Marin, M.; Vogel, S. N.; et  al., 2012a, A\&A 537, 52
\bibitem{} Eckart, A.; Britzen, S.; Horrobin, M.; et al.; 2012b,
Proceedings of Nuclei of Seyfert galaxies and QSOs - Central engine \& conditions of star formation 
(Seyfert 2012). 6-8 November, 2012. Max-Planck-Insitut f\"ur Radioastronomie (MPIfR), Bonn, Germany
\bibitem{} Eckart, A., Sch\"odel, R.; Garcia-Marin, M.; Witzel, G.; et al. 2008, A\&A 492, 337
\bibitem{} Eckart, A.; Baganoff, F. K.; Morris, M.; et al., 2004, A\&A 427, 1
\bibitem{} Eckart, A.; Moultaka, J.; Viehmann, T.; Straubmeier, C.; Mouawad, N.; 2004, ApJ 602, 760	
\bibitem{} Eckart, A.; Genzel, R., 1997, MNRAS 284, 576	
\bibitem{} Edelson, R.; Gelbord, J. M.; Horne, K.; et al., 2015, ApJ 806, 129
\bibitem{} Ferrarese, L.; Merritt, D., 2000, ApJ 539, L9
\bibitem{} Gebhardt, K., Bender, R. et al.; 2000, ApJ 539, L13
\bibitem{} Garcia-Burillo, S., \& Combes, F.; Journal of Physics: Conference Series, Volume 372, Issue 1, id. 012050 (2012)
\bibitem{} Garcia-Marin, M.; Eckart, A.; Weiss, A.; et al.; 2011, ApJ 738, 158
\bibitem{} Gillessen, S., Genzel, R.; Fritz, T. K.; Quataert, E.; et al. 2012, Nature 481, 51
\bibitem{} Higginbottom, N.; Knigge, C.; Long, K.S.; et al., 2013, MNRAS 436, 1390
\bibitem{} Horne, K., \& Peterson, B.; 2004, PASP 116, 465
\bibitem{} Jalali, B.; Pelupessy, F. I.; Eckart, A.; Portegies Zwart, S.; et al., 2014, MNRAS 444, 1205
\bibitem{} Kormendy, J. and Kennicutt, Jr., R. C., Ann. Rev. Astron. Astrophys., 42 , 603-683, 2004.
\bibitem{} Kormendy, J., Bender, R., and Cornell, M. E., Nature, 469 , 374-376, 2011.
\bibitem{} Kunneriath, D.; Eckart, A.; Vogel, S. N.; Teuben, P.; et al., 2012, A\&A 538, 127
\bibitem{} Lee, J.W.; Lee, S.-S.; et al., 2016, A\&A 592, L10
\bibitem{} Li, Ya-Ping; Yuan, Feng; Wang, Q. Daniel; 2017, MNRAS 468, 2552
\bibitem{} Lister, M. L.; Aller, M. F.; Aller, H. D.; et al.; 2016, AJ 152, 12
\bibitem{} Liu, Z.; Yuan, W.; Lu, Y.; Carrera, F.J.; et al.; 2016, MNRAS 463, 684
\bibitem{} L\"utzgendorf, N.; Kissler-Patig, M.; et al.; 2013, A\&A 555, 26
\bibitem{} Micic, M.; Martinovic, N.; Sinha, M., 2016, MNRAS 461, 3322
\bibitem{} Narayan, R., Mahadevan, R., Grindlay, J. E., et al.; 1998, ApJ, 492, 554
\bibitem{} Neilsen, J., Nowak, M. A., Gammie, C., et al.  2013, ApJ, 774, 42
\bibitem{} Nowak, M. A., Neilsen, J., Markoff, S. B., et al.  2012, ApJ, 759, 95
\bibitem{} Magorrian J., Tremaine S., Richstone D., et al, 1998, The Astronomical Journal, 115, 2285
\bibitem{} Marin, F.; Tamborra, F.; 2014, AdSpR 54, 1458
\bibitem{} Marrone, D.P. 2006, Ph.D. Thesis, Harvard Univ.
\bibitem{} Marrone, D.P., Moran, J.M., Zhao, J.-H., \& Rao, R. 2006a, ApJ, 640, 308
\bibitem{} Marrone, D.P., Moran, J.M., Zhao, J.-H., \& Rao, R. 2006b, JPhCS, 54, 354
\bibitem{} Morse J. A., Wilson A. S., Elvis M., Weaver K. A., 1995, ApJ, 439, 121
\bibitem{} Moser, L.; Sanchez-Monge, A.; Eckart, A.; Requena-Torres, M.A.; Garcia-Marin, M.; et al.  2016arXiv160300801M	
\bibitem{} Moscibrodzka, M.; Falcke, H., 2013, A\&A 559, L3	
\bibitem{} Moscibrodzka, M.; Gammie, C.F.; et al., 2009, ApJ 706, 497
\bibitem{} Mossoux, E.; Grosso, N.; Bushouse, H.; Eckart, A.; Yusef-Zadeh, F.; et al., 2016, A\&A 589, 116
\bibitem{} Matthews, J.H.; Knigge, C.; Long, K. S.; et al., 2016, MNRAS 458, 293
\bibitem{} Narayan, R.; Sadowski, A.; Penna, R.F.; Kulkarni, A.K., 2012, MNRAS 426, 3241
\bibitem{} Narayan, R.; Mahadevan, R., 1998, ApJ 492, 554
\bibitem{} Narayan, R.; Garcia, M.R.; McClintock, J.E., 1997, ApJ 478, L79
\bibitem{} Nayakshin, S.; Cuadra, J., 2005, A\&A 437, 437
\bibitem{} Nayakshin, S.; Cuadra, J.; Springel, V., 2007, MNRAS 379, 21	
\bibitem{} Ogle P. M., Marshall H. L., Lee J. C., Canizares C. R., 2000, ApJ, 545, L81
\bibitem{} Ortiz-Leon, G.N.; Johnson, M.D.; Doeleman, S.S.; et al., 2016, ApJ 824, 40O	
\bibitem{} Peissker, F.; Valencia-S., M.; Eckart, A.;  Parsa, M.; Zajacek, M.; \& Shahzamanian, B.; in prep. 2017
\bibitem{} Pfuhl, O.; Gillessen, S.; Eisenhauer, F.; Genzel, R., 2015, ApJ 798, 111	
\bibitem{} Penston M. V. et al., 1990, A\&A, 236, 53
\bibitem{} Porquet, D., Predehl, P., Aschenbach, B., et al.  2003, A\&A, 407, L17
\bibitem{} Porquet, D., Grosso, N., Predehl, P., et al. 2008, A\&A, 488, 549
\bibitem{} Peterson B.M. et al., 2013, ApJ, 779, 109
\bibitem{} Peterson, B.M.; 1993, PASP 105, 247	
\bibitem{} Peterson, B.M.; Ferrarese, L.; et al.; 2004, ApJ 613, 682
\bibitem{} Rashed, Y.E.; Zuther, J.; Eckart, A.; et al.; 2013, A\&A 558, 5
\bibitem{} Rashed, Y. E.; Eckart, A.; Valencia-S., M.; Garcia-Marin, M.; et al.; 2015, MNRAS 454, 2918
\bibitem{} Rauch, C.; Ros, E.; Krichbaum, T. P.; Eckart, A.; Zensus, J. A.; Shahzamanian, B.; Muzic, K.  2016, A\&A 587, 37
\bibitem{} Schartmann, M.; Ballone, A.; Burkert, A.; Gillessen, S.; et al.i,  2015, ApJ 811, 155
\bibitem{} Shahzamanian, B., Eckart, A., Zajacek, M., Valencia-S. , M. , et al., 2016, A\&A 593, 131
\bibitem{} Shahzamanian, B.; Eckart, A.; Valencia-S., M.; Witzel, G.; et al.  2015, A\&A 576, 20
\bibitem{} Smajic, S.; Moser, L.; Eckart, A.; Busch, G.; Combes, F. et al., 2015, A\&A 583, 104	
\bibitem{} Sobacchi, E.; S., Mattia C.; Stamerra, A.,  2017, MNRAS 465, 161
\bibitem{} Subroweit, M., Garcia-Marin, M., Eckart, A., Borkar, A., Valencia-S., M., Witzel, G., Shahzamanian, B. , and Straubmeier, C., A\&A, submitted
\bibitem{} Tremaine, S.; Gebhardt, K.; et al.; 2002, ApJ 574, 740
\bibitem{} Urry C. M., Padovani, P., 1995, Publications of the Astronomical Society of the Pacific, 107, 803
Mossoux, E.; Grosso, N.; Bushouse, H.; Eckart, A.; Yusef-Zadeh, F.; 2016, On-line Data Catalog: J/A+A/589/A116
\bibitem{} Valencia-S., Eckart, A., et al. 2015, ApJ 800, 125
\bibitem{} Witzel, G., Eckart, A.; Bremer, M.; Zamaninasab, M.; et al., 2012, ApJS 203 18
\bibitem{} Witzel, G.; Ghez, A. M.; Morris, M. R.; et al., 2014, ApJ 796, L8
\bibitem{} Yang, H.-Y. K.; Reynolds, C. S., 2016, ApJ 829, 90	
\bibitem{} Yuan, F., Quataert, E., \& Narayan, R. 2003, ApJ, 598, 301
\bibitem{} Yuan, Feng; Lin, Jun; Wu, Kinwah; Ho, Luis C.; 2009, MNRAS 395, 2183
\bibitem{} Yusef-Zadeh, F.; Bushouse, H.; Dowell, C. D.; et al., 2006, ApJ 644, 198
\bibitem{} Yusef-Zadeh, F.; Roberts, D.; Wardle, M.; Heinke, C. O.; Bower, G. C., 2006, ApJ 650, 189
\bibitem{} Yusef-Zadeh, F.; Wardle, M.; Sch\"odel, R.; et al., 2016, ApJ 819, 60
\bibitem{} Yusef-Zadeh, F.; Wardle, M.; Sewilo, M.; 2015, ApJ 808, 97	
\bibitem{} Zajacek, M.; Karas, V.; Eckart, A., 2014, A\&A 565, 17
\bibitem{} Zajacek, M.; Eckart, A.; Karas, V.; Kunneriath, D.; Shahzamanian, B.; Sabha, N.; Muzic, K.; Valencia-S., M.; 2016, MNRAS 455, 1257
\end{thebibliography}
\end{document}